\documentclass[twocolumn,showpacs,preprintnumbers,amsmath,amssymb]{revtex4}
\usepackage{graphicx}% Include figure files
\usepackage{psfig}% Include figure files
\usepackage{dcolumn}% Align table columns on decimal point
\usepackage{bm}% bold math
\usepackage{multirow}
\newcommand{\elem}[2]{\ensuremath{{}^{#1}}#2}

\begin{document}

\title{Medium polarization isotopic effects on nuclear binding energies}

\author{S.Baroni$^{a,b}$, F.Barranco$^{c}$, P.F.Bortignon$^{a,b}$, R.A.Broglia$^{a,b,d}$,
        G.Col\`o $^{a,b}$, E.Vigezzi$^{b}$} 
\affiliation{
$^a$ Dipartimento di Fisica, Universit\`a degli Studi di Milano,via Celoria 16, 20133 Milano, Italy.\\
$^b$ INFN, Sezione di Milano, via Celoria 16, 20133 Milano, Italy.\\
$^c$ Departamento de Fisica Aplicada III, Escuela Superior de Ingenieros, c
Camino de los Descubrimientos s/n,  
  41029 Sevilla, Spain.\\
$^d$ The Niels Bohr Institute, University of Copenhagen, Blegdamsvej 17, 2100 Copenhagen \O, Denmark.}

\date{\today}


\begin{abstract}

There exist several effective interactions whose parameters are fitted to force mean field predictions 
to reproduce experimental findings of finite nuclei and calculated properties of infinite nuclear matter.
Exploiting this tecnique one can give a good description of nuclear binding energies. 
We present evidence that further progress can be made
by taking into account medium polarization 
effects associated with surface and pairing vibrations.
\end{abstract}

\maketitle

Mean field theory is one of the most useful approximations in all of physics and chemistry. 
In it we replace the many-particle Schr\"odinger equation by the single particle 
Schr\"odinger equation. 
What started out as a problem too complicated to solve for all but the smallest systems becomes quite manageable when 
one deals with one particle at a time (see e.g. \cite{Ben.ea:2003}).
Mass formulae based on mean field theory can achieve good agreement with experimental findings. When phenomenological
corrections are added to take into account selected correlation effects (e.g. rotational energy correction, a Wigner term
to take into account the proton-neutron interaction in nuclei with N nearly equal to Z) one can
fit the experimental masses with a rms deviation of about 0.6 MeV.

In the present paper we study some of the consequences zero point fluctuations associated with surface and pairing
vibrations have on nuclear masses (in connection with the first subject see ref \cite{Ben.ea:2005}).

\section{Details of the calculation}\label{sec:calc}

In what follows we concentrate on a set of 121 spherical even-even nuclei (ranging from $A=16$ to $A=214$).

We started from a mean field calculated in the HF approximation, according 
to the procedure followed by Goriely et al. \cite{Gor.ea:2001}. In particular, we 
made use of the Skyrme-type MSk7 interaction, adding 
a phenomenological Wigner term in order to take into account neutron-proton correlations
 in nuclei
having nearly equal number of neutrons and protons.
The pairing channel was treated in the BCS approximation using a zero-range interaction 
\begin{equation}\label{eq:art2 4}
  V_{pair}(\vec{r},\vec{r}')=V_0\delta(\vec{r}-\vec{r}')
\end{equation}
in which we allow the pairing strength parameter $V_0$ to be different 
for neutrons and protons and
with an energy cutoff
corresponding to one major shell ($E=41 \cdot A^{-1/3}$ MeV).
HFBCS solutions were computed in a box of 16 fm radius, making use of a radial mesh of 0.1 fm.

We then focused on the effect of correlations on the binding energies.
The contribution associated with surface vibrations
has been calculated in the quasiparticle random phase approximation
(QRPA), according to the expression 
\begin{equation}\label{eq:art2 6}
    E^{corr}_{QRPA}=-(2\lambda + 1)\sum_{\lambda,n}\hbar\omega_{\lambda}(n)
                   \sum_{ki}\left| Y_{ki}^{\lambda}(n) \right|^{2},
\end{equation}
where 
$\hbar\omega_{\lambda}(n)$ are the energies of the n-th QRPA phonons 
of multipolarity $\lambda$, $Y_{ki}^{\lambda}(n)$ are
the backwardsgoing amplitudes (associated with the corresponding vibrational modes of energy $\hbar\omega_{\lambda}(n)$) and
the sums over $k$ and $i$ run on QRPA quasiparticle indices
(on RPA particle and hole indices for nuclei which are not superfluid).
We don't consider here other important renormalization effects, 
e.g. self-energy effects, which affect the value of the  effective mass ($m^* = 1.05 m$ in the case of
the MSk7 interaction) and 
the position of single-particle levels and level densities. 
In QRPA calculations, 
using the same interactions  adopted in the HF+BCS mean field,  
$2^+$ and $3^-$ vibrations have been calculated using a space dimension 
ranging from 40 MeV for the heaviest isotopes to 64 MeV for the lightest ones,
exhausting the energy-weighted sum rule.
Only the  most collective low-lying QRPA phonons have been considered for the 
computation of correlation energies.
The selection of the most collective phonons is achieved taking into account only
those QRPA states with a reduced transition probability 
greater than 2\% of non-energy weighted sum rule, see also ref. \cite{Bar.ea:2004}; 
for what concerns the individualization of the low-lying states, the
fragmentation of the QRPA response imposes a case by case analysis, but a simple rule
came out: quadrupolar (octupolar) states up to 7 MeV (5 MeV) contributed to the calculation of correlation energies,
except for some nuclei (namely, Ar, Ca, Ti, Zr) for which we took into account quadrupolar (octupolar)
states up to 10 MeV (7 MeV).

Pairing vibrations  (see \cite{Bro.ea:1973},\cite{Bri.Bro:2005}) were treated in RPA. 
While it is true that well developed pairing vibrational bands based on doubly closed nuclei  
have been observed (e.g. in the case of \elem{208}{Pb}, see \cite{Bro.ea:1973}),
one expects pairing collectivity to be important also for systems with few nucleons outside closed shells.
However, the correlation energy associated with pairing vibrations is calculated only 
on shell closures, while the correlations associated with pairing in open shell nuclei have been taken into account
by means of the BCS approach.
In any case, the above is an approximation which, if needed, can be removed from the calculation scheme.
Single particle levels have been computed using 
a Woods-Saxon potential with standard parametrization \cite{Boh.Mo:1969}. 
A separable pairing interaction with constant matrix elements has been used to compute 
monopole pairing vibration contributions.
In all calculations we have kept the contribution of only the lowest
($n = 1$) pair addition and pair subtraction modes, in keeping with the fact that,
as a rule, the $n\neq 1$ modes are much less collective.
The values of the associated pairing strengths in a specific nucleus have been computed 
as in ref. \cite{Bar.ea:2004} (see note \cite{mat.ele}).
For neutron pairing vibrations the pairing strength has been computed in double closed shell nuclei
and the resulting value has been used for isotopes with the same magic number of neutrons
(\elem{16}{O},\elem{40}{Ca},\elem{48}{Ca},\elem{132}{Sn},\elem{208}{Pb} for N=8,20,28,82,126 respectively), while 
for N=50 the value has been computed in \elem{90}{Zr}.
For isotopic chains with magic atomic number, the correlation energy due to proton pairing vibrations has been calculated 
only for selected isotopes 
(\elem{16-22}{O},\elem{40}{Ca},\elem{48}{Ca},\elem{110}{Sn},\elem{128}{Sn},\elem{208}{Pb} for Z=8,20,50,82)
and then interpolated for the rest of the chain.

\section{Results}\label{sec:results}

\subsection{Mean field}\label{sub:MF}
\begin{figure}[!h]
\centerline{\psfig{file=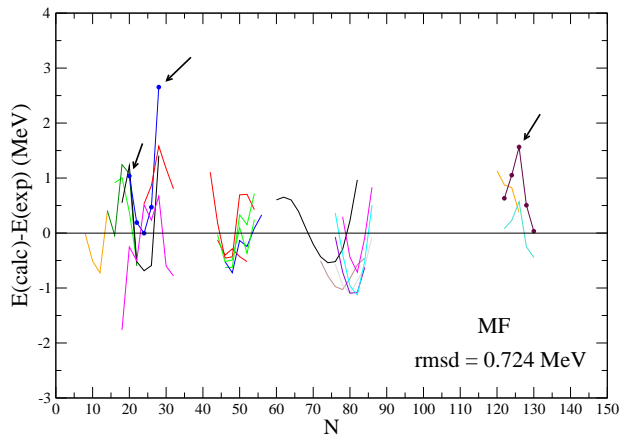,width=9.5cm,angle=-90}}
\caption{Deviations of computed mean field ground state energies from experimental values as a function of neutron number.
The straight horizontal line is the $E^{calc}_j - E^{exp}_j = 0$ line. Nuclei above this line are underbound.
Isotopic chains are connected by solid lines. The full dots highlight calcium and lead isotopes, commented in the text.
Figures come in color in the online version.}
\label{fig:art2 10}
\end{figure}

Fig. \ref{fig:art2 10} shows the deviations of computed mean field (MF) ground state energies $E^{calc}$
from experimental values $E^{exp}$ (\cite{Aud.Wap:1995},\cite{note:BE}).
The r.m.s. deviation 
\begin{equation}\label{eq:art2 8}
  rmsd=\sqrt{\frac{1}{N}\sum_j \left( ( E^{calc}_j - E^{exp}_j) \right)^2}
\end{equation}
we obtain for this set of 121 nuclei is 0.724 MeV. 
Note that the parameters of the Skyrme force used are slightly different from the MSk7 developed by Goriely et al. in \cite{Gor.ea:2001}.
We started with the MSk7 parametrization and, with the aid of a linear refit (which will be discussed below),
we fitted the Skyrme parameters to the experimental binding energies of this set of 121 isotopes. The 
resulting parametrization was used as a starting point of the present work \cite{note: 263}.

The shell structure dependence  of the results is evident from Fig. \ref{fig:art2 10}. At variance with the results 
reported in ref. \cite{Ben.ea:2005} obtained making use of the SLy4 Skyrme parametrization,
magic number nuclei are not, in our calculations, regularly overbound with respect to 
their neighbours. In fact, \elem{40}{Ca},\elem{48}{Ca} and \elem{208}{Pb} (indicated by the arrows in the plot) 
are examples of magic nuclei which are underbound with respect to near isotopes, while for N=132
some opposite cases are observed.

\subsection{Correlation energies}\label{sub:corr}

\begin{figure}[!h]
\centerline{\psfig{file=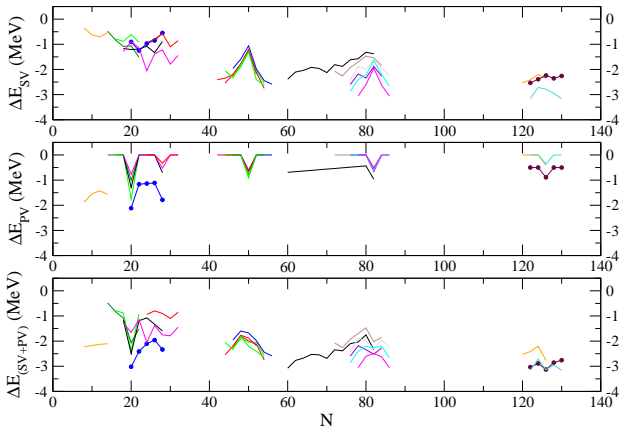,width=9.5cm,angle=-90}}
\caption{Correlation energies associated to surface vibrations (SV,upper panel), pairing vibrations (PV,middle
panel) and the sum of them (SV+PV,lower panel). Isotopic chains are connected by solid lines. All correlation energies are negative.
The full dots highlight calcium and lead isotopes, as in Fig.\ref{fig:art2 10}.}
\label{fig:art2 20}
\end{figure}

Surface and pairing correlation energies are plotted in Fig. \ref{fig:art2 20}.
Low-lying surface vibrations show, as expected, a marked collectivity for all the isotopes calculated. 
These corrections tend to bind
magic nuclei less than their neighbours (compare Fig. \ref{fig:art2 20} for the neutron shell closures
at $N=20,28,50,82$). On the other hand pairing vibrations contributions show 
the opposite
behaviour, providing magic nuclei with an important extra binding energy. 
The sum of the two contributions is displayed in the lower panel
of Fig. \ref{fig:art2 20}. The differences which can be found between the results presented in ref. \cite{Bar.ea:2004}
and those presented in this work
arise from the inclusion in the previous paper of quadrupole and 
hexadecapole pairing vibrations and from the fixing of some computational errors.

After adding surface and pairing correlations to mean field binding energies, a refit of the Skyrme
parameters is needed in order to minimize the rms deviation (\ref{eq:art2 8}). After a first refit
of the parameters one should recompute all nuclei in the mean field approximation and also recompute the 
correlations. Since a full fit of the Skyrme interaction to the experimental binding energies is too costly, we will
follow here the procedure of ref. \cite{Ber.ea:2005}, readjusting 
perturbatively the parameters of the force \cite{note:RF}.
This implies that the new interaction will be, in the parameters space, rather similar to the original MF interaction.
Thus, QRPA correlations are expected to change little under the refitting process. This is the reason why
we leave unchanged the correlation energy contributions coming from QRPA.

In the case of surface vibrations, we have compared our results with those recently obtained
by  Bender et al. \cite{Ben.ea:2005,{Ber:pc}}.
We started from their mean field binding energies 
computed with the SLy4 Skyrme interaction within the HF+BCS framework 
for the 121 isotopes studied in the present paper; 
these binding energies display a rms deviation from the experimental values
of 2.45 (1.54) MeV before (after) a refit procedure is performed (see below for a brief 
explanation or \cite{Ber.ea:2005} for a more detailed
one). If one computes the correlation energies obtained projecting on good
angular momentum ($J=0$), as the authors of \cite{Ben.ea:2005} did, and adds 
these corrections to the
mean field binding energies one obtains (after refitting) a rms deviation of 1.45 MeV. 
If instead one adds the surface correlation energies showed in the middle panel of
Fig. \ref{fig:art2 20}, one obtains a rms deviation of 1.46 MeV.

In what follows we give a brief description of the perturbative refit performed (see ref. \cite{Ber.ea:2005} for a detailed discussion).
The mean field Skyrme energy functional is decomposed into a sum of integrals $f_i$, each of which is weighted by some linear combination
$c_i$ of the Skyrme parameters
\begin{equation}\label{eq:art2 30}
  E=\sum_{i=1}^{10} c_i f_i.
\end{equation}
These integrals $f_i$ and the mean field binding energies are sufficient to perform the refit of the
interaction. Since certain linear combinations of the Skyrme parameters are not determined by nuclear masses, one should not include them
in the fit. According to \cite{Ber.ea:2005} only four combinations are well fixed by the binding energies.
The refit procedure is linear in these integrals $f_i$ and one can vary the $c_i$ coefficients in order to minimize 
the rms deviation between theoretical and experimental binding energies. We tested the linearity of the procedure
recomputing the binding energies with the new parameters and comparing these energies to those obtained exploiting
the linearity, finding excellent agreement.

\begin{figure}[!h]
\centerline{\psfig{file=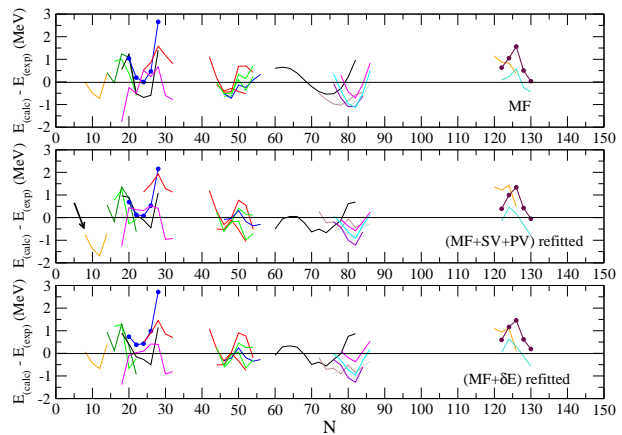,width=9.5cm,angle=-90}}
\caption{Deviations of binding energies for mean field MF (upper panel), mean field MF corrected by QRPA correlations 
after the refit to experimental binding energies (middle panel) and mean field MF plus correlation energy fluctuations $\delta E$
(discussed below in the text)
after the refit to experimental binding energies. Full dots highlight, as in previous figures, calcium and lead isotopic chains.
Isotopic chains are connected by solid lines.}
\label{fig:art2 50}
\end{figure}

The results obtained 
after performing the refitting procedure are shown in Fig. \ref{fig:art2 50}.
The rms deviation for the refit of mean field plus surface and pairing correlations
is 0.714 (middle panel of Fig. \ref{fig:art2 50}), very close to the 
mean field value (0.724 MeV).
It can be seen that adding the negative correlations 
has increased the deviation of the calculated masses from 
the experimental data for a few specific isotopic chains,
which are now  systematically overbound (cf. for example the oxygen
isotopes, indicated by an arrow in the middle panel of Fig. \ref{fig:art2 50}), notwithstanding the refitting procedure.

To gain perspective concerning this point, it is worth stepping back and looking at 
what happens when the ground state correlation energies are added to the isotopic chains studied, 
without carrying out the refitting procedure.
It is seen that, adding the correlation energy contributions, the deviation between theory and experiment 
for some of the nuclear chains studied displays a flatter isotopic dependence. An index of this flattening is provided by 
the rms deviation calculated with respect 
to the average deviation of the single isotopic chain. In other words, for the $k-th$ isotopic chain
\begin{equation}\label{eq:art2 10}
  \sigma_k=\sqrt{\frac{1}{N_k}\sum_i^{N_k} \left( ( E^{calc}_i - E^{exp}_i) - \alpha_k \right)^2}
\end{equation}
where $\alpha_k$ is the average deviation
\begin{equation}\label{eq:art2 20}
  \alpha_k=\frac{\sum_i^{N_k} \left( E^{calc}_i - E^{exp}_i\right)}{N_k}.
\end{equation} 

The values of $\sigma_k$ for all the 21 isotopic chains studied
are displayed in Table \ref{table:art2 10}.
It can be seen that in many cases (13  out of 21) $\sigma_k$ 
decreases compared to the mean field value, mostly due to the effect of
pairing vibrations. The effect of surface vibrations alone is to
increase $\sigma_k$ compared to the mean field value in 16 chains out of 21.
The total value of $\sigma_k$ for all the isotopic chains, defined by the expression
\begin{equation}\label{eq:art2 22}
  \sigma^{total}=\sqrt{\frac{\sum_k N_k \sigma_k^2}{\sum_k N_k}}
\end{equation}
decreases by about 6\% from 0.532 MeV for the MF calculation 
to the corrected mean field (0.502 MeV).

In keeping with this finding, we have added to mean field
only the fluctuations of the QRPA correlations. That is, we added to each isotope 
the difference $\delta E$ between its correlation energy
and the average correlation energy of the isotopic chain it belongs to. This procedure clearly cancels out 
the information related
to the absolute value of the computed correlation energies, 
but keeps the, arguably, more important information concerning the isotopic
dependence of these energies. 
In this context, we can remark that while the absolute value of the correlation
energies associated with collective surface vibrations depends on the choice of the 
energy of the phonons which contribute to these correlation energies, 
the isotopic dependence is much more stable. This is because the properties
of giant resonance modes change little along a given isotopic chain.
Consequently, adding the fluctuations $\delta E$ to MF 
selects the most stable fraction of the contributions associated with
ground state correlation energies.
The results obtained after refitting are shown in the lower panel of Fig. \ref{fig:art2 50}. 
The rms deviation is now 0.691 MeV.

\renewcommand{\multirowsetup}{\centering}
\begin{table}[!h]
\centering
\begin{tabular}{|c|c|c|c|c|}
%  \multicolumn{2}{c}{} & \multicolumn{3}{c}{rms errors (MeV)} \\ 
  \cline{2-5}
  \multicolumn{1}{c|}{} & \multicolumn{1}{c|}{\#} & \multicolumn{1}{c|}{$\sigma_k${\tiny (MF)}} 
    & \multicolumn{1}{c|}{$\sigma_k${\tiny (MF+SV)}} & \multicolumn{1}{c|}{$\sigma_k${\tiny (MF+SV+PV($0^+$))}} \\ \hline
  {\bf O}  (16 - 22) &  4 & {\bf 0.430} & 0.524 & {\bf 0.436} \\ \hline 
  {\bf Si} (30 - 36) &  4 & {\bf 0.603} & 0.642 & {\bf 0.939} \\ \hline
  {\bf S} (30-38) & 5 & {\bf 0.688} & 0.850 & {\bf 0.803} \\ \hline 
  {\bf Ar} (36 - 46) &  6 & {\bf 0.871} & 0.946 & {\bf 0.639} \\ \hline
  {\bf Ca} (40 - 48) &  5 & {\bf 0.959} & 1.153 & {\bf 0.941} \\ \hline 
  {\bf Ti} (40 - 54) &  8 & {\bf 0.741} & 0.751 & {\bf 0.621}  \\ \hline
  {\bf Fe} (50 - 58) &  5 & {\bf 0.358} & 0.468 & {\bf 0.350} \\ \hline 
  {\bf Se} (78 - 86) &  5 & {\bf 0.136} & 0.414 & {\bf 0.250} \\ \hline
  {\bf Kr} (80 - 90) &  6 & {\bf 0.296} & 0.547 & {\bf 0.298} \\ \hline
  {\bf Sr} (80 - 92) &  7 & {\bf 0.548} & 0.693 & {\bf 0.519} \\ \hline 
  {\bf Zr} (86 - 94) &  5 & {\bf 0.535} & 0.662 & {\bf 0.461} \\ \hline  
  {\bf Mo} (88-98) &  6 & {\bf 0.352} & 0.429 & {\bf 0.208}  \\ \hline 
  {\bf Sn} (110-132) & 12 & {\bf 0.498} & 0.551 & {\bf 0.473} \\ \hline 
  {\bf Te} (124 - 136) &  7 & {\bf 0.211} & 0.316 & {\bf 0.258}  \\ \hline 
  {\bf Xe} (130- 140) &  6 & {\bf 0.333} & 0.278 & {\bf 0.301} \\ \hline
  {\bf Ba} (132-140) &  5 & {\bf 0.373} & 0.250 & {\bf 0.354}  \\ \hline
  {\bf Ce}  (134 - 144) &  6 & {\bf 0.612} & 0.306 & {\bf 0.407}  \\ \hline  
  {\bf Sm} (140-148) &  5 & {\bf 0.543} & 0.263 & {\bf 0.333}  \\ \hline
  {\bf Hg} (200-206) &  4 & {\bf 0.270} & 0.244 & {\bf 0.416} \\ \hline
  {\bf Pb} (204-212) &  5 & {\bf 0.518} & 0.541 & {\bf 0.417} \\ \hline 
  {\bf Po} (206-214) &  5 & {\bf 0.359} & 0.502 & {\bf 0.416}  \\ \hline
  {$\sigma_{total}$} & {121} & { \bf 0.532} 
    & {0.591} & {{\bf 0.502}} \\ \hline

\end{tabular}
\caption{Rms deviations $\sigma_k$ of calculated ground state energy 
from experimental values are shown.
These deviations are calculated with respect to the average deviation for each isotopic chain (see eqs. \ref{eq:art2 10} and
\ref{eq:art2 20}). 
The Table displays: the mass number range (1st column), the 
number of isotopes calculated (2nd), the mean field rmsd $\sigma_k$ (3rd), 
the rmsd obtained adding the contribution of surface vibrations to 
the mean field ground state energy (4th), the rmsd obtained adding the contributions from pairing vibrations as well (5th). 
The last row shows $\sigma_{total}$ for all the isotopic chains (see eq. \ref{eq:art2 22}). All
numbers are in MeV.}
\label{table:art2 10}
\end{table}

\subsection{Further improvements}\label{sub:furth}

\begin{figure}[!h]
\centerline{\psfig{file=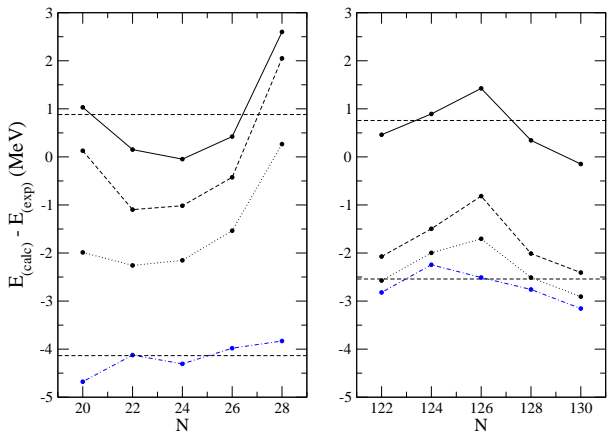,width=9.5cm,angle=-90}}
\caption{The effect of adding surface and pairing correlation energies to mean field binding energies are plotted for Ca (left)
and Pb (right) isotopic chains. Solid lines represents mean field calculation results (MF), shown for all
the 121 isotopes in Fig. \ref{fig:art2 10}. Dashed lines are obtained adding the surface correlations of the
upper panel of Fig. \ref{fig:art2 20} to the mean field binding energies (MF+SV), dotted lines
are obtained adding the monopole pairing correlations of the middle panel of Fig. \ref{fig:art2 20} as well (MF+SV+PV($0^+$)),
and  dot-dashed lines represent the results obtained adding quadrupole pairing contributions as well (MF+SV+PV($0^+$,$2^+$)).
The dashed horizontal lines indicate the average values of the first and the last curves, in order to better judge the improvement
gained going from MF to MF+SV+PV($0^+$,$2^+$).}
\label{fig:art2 30}
\end{figure}

\begin{figure}[!h]
\centerline{\psfig{file=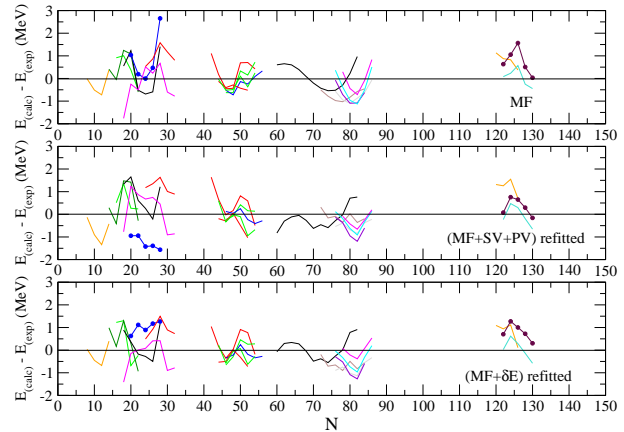,width=9.5cm,angle=-90}}
\caption{The same for Fig. \ref{fig:art2 50}, but with the inclusion of quadrupole pairing vibrational correlations
for calcium and lead isotopes.}
\label{fig:art2 60}
\end{figure}

Observing Table \ref{table:art2 10} one can realize that for some isotopic chains (for example calcium, tin and 
sulphur)
$\sigma_k$ mantains nearly the same value or increases after adding surface and pairing vibrational correlation energies.
The results shown in the Table arose from the application of the prescription described in section \ref{sec:calc}.
This prescription is the same for all nuclei we included in this study, and this is in the spirit of a systematic study
of the effect of these correlations to nuclear masses.
Nevertheless it is worth looking at what we can gain by studying certain chains in more detail, in order to learn something
that could be useful to improve our general prescription.
If we focus on calcium and lead isotopes, for example, we know that
in these nuclei well developed quadrupole pairing vibrations have been observed \cite{Bro.ea:1973}. This lead us to
include in these isotopes, besides the monopole pairing contribution, also the quadrupole one.
The corresponding results are shown in Fig. \ref{fig:art2 30} and are summarized in Table \ref{table:art2 20}. The improvement is 
evident.
In the middle panel of Fig. \ref{fig:art2 60} we show the results obtained 
performing the refit on all 121 nuclei, adding quadrupole  pairing vibrations for calcium and lead;
in the lower panel, we show the results obtained adding only the flucutations $\delta E$. In the latter case, 
the rms deviation for all 121 isotopes is 0.667 MeV (that is, a reduction of the rmsd from MF of nearly 8\%).

Analyses like this one could be pursued in other regions of the chart of nuclides, and it
may be possible to find out some general rules that one can include in a general prescription.

\renewcommand{\multirowsetup}{\centering}
\begin{table}[!h]
\centering
\begin{tabular}{|c|c|c|c|c|}
  \cline{2-5}
  \multicolumn{1}{c|}{} & \multicolumn{1}{c|}{$\sigma_k${\tiny (MF)}} 
    & \multicolumn{1}{c|}{$\sigma_k${\tiny (MF+SV)}} & \multicolumn{1}{c|}{$\sigma_k${\tiny (MF+SV+PV($0^+$))}}
    & \multicolumn{1}{c|}{$\sigma_k${\tiny (MF+SV+PV($0^+$,$2^+$))}} \\ \hline
  {\bf Ca} &  {\bf 0.959} & 1.153 & 0.941 & {\bf 0.308} \\ \hline 
  {\bf Pb} &  {\bf 0.518} & 0.541 & 0.417 & {\bf 0.296} \\ \hline 
\end{tabular}
\caption{Rms deviations $\sigma_k$ for calcium and lead after the
addition of the correlation energy contributions described in the text are shown.
These deviations are calculated with respect to the average deviation for each isotopic chain (see eqs. \ref{eq:art2 10} and
\ref{eq:art2 20}). 
The Table columns display the mean field rmsd (2nd column), the rmsd for mean field plus surface correlation (3rd), 
the rmsd obtained adding monopole pairing contributions as well (4th) and the rmsd when also quadrupole pairing vibrational
contributions are included. All numbers are in MeV.}
\label{table:art2 20}
\end{table}

\subsection{Influence of the various contributions}

It is worth looking at what happens when we add to the mean field binding energy only some of the contributions 
to correlation energies: data are shown in Table \ref{table:art2 30}. 
It is evident, if one looks at
the rms deviations obtained adding the fluctuations $\delta E$ of the correlation energies
(last column of Table \ref{table:art2 30}),
that the octupolar surface correlation energy has a small influence on the results.
This is observed
in all the three cases shown in Table \ref{table:art2 30}. 

Comparing the first and the third rows of this Table (and analogously the second and the fourth), 
one can realize that the quadrupole contribution to pairing
vibration correlations has a large influence on the results. In particular, adding this contribution to all isotopes, 
rather than adding them only to calcium and lead isotopes, leads
to a worse rms deviation (compare the first and the fifth rows of Table \ref{table:art2 30}).

The values of $\sigma_k$ (see eq. \ref{eq:art2 10}) for the various prescriptions are shown in Table \ref{table:art2 50}.
It is seen that the results in Table III and IV follow the same trends, the best agreement with experiment
being obtained in the case $(e)$, namely including quadrupole pairing vibrations
for calcium and lead isotopes.

\renewcommand{\multirowsetup}{\centering}
\begin{table}[!h]
\centering
\begin{tabular}{c|cc|cc|c|c|c}
  \cline{2-7}
  \multicolumn{1}{c|}{} & \multicolumn{2}{c|}{SV} & \multicolumn{2}{c|}{PV} 
    & \multicolumn{1}{c|}{rmsd} & \multicolumn{1}{c|}{rmsd} & \multicolumn{1}{c}{} \\ % \hline
  \multicolumn{1}{c|}{} & \multicolumn{1}{c}{$2^+$} & \multicolumn{1}{c|}{$3^-$} &
    \multicolumn{1}{c}{$0^+$} & \multicolumn{1}{c|}{$2^+$}
    & \multicolumn{1}{c|}{MF+SV+PV} & \multicolumn{1}{c|}{MF+$\delta E$} & \multicolumn{1}{c}{} \\  \cline{2-7}
  \multirow{8}{10mm}{MF+}
    & X & X & X & X & 0.935 & 0.788 & \emph{(a)} \\
    & X &   & X & X & 0.942 & 0.817 & \emph{(b)} \\ \cline{2-7}
    & X & X & X &   & 0.714 & 0.691 & \emph{(c)} \\
    & X &   & X &   & 0.702 & 0.694 & \emph{(d)} \\ \cline{2-7}
    & X & X & X & only Ca & 0.729 & 0.667 & \emph{(e)} \\ 
    &   &   &   & and Pb     &       &       \\
    & X &   & X & only Ca & 1.225 & 0.677 & \emph{(f)} \\ 
    &   &   &   & and Pb     &       &       \\ \cline{2-7}
\end{tabular}
\caption{Rms deviations obtained after adding some of the contributions to correlation energies to the 
mean field ground state of the 121 isotopes.
The Table columns display the contributions taken into account in each case  (first four columns), 
the rmsd (obtained after the refit) for mean field plus these correlations  (5th column) and 
the rmsd (obtained after the refit) for mean field plus the fluctuations $\delta E$ of these correlations (6th).
All numbers are in MeV. The values should be compared to the mean field rmsd of 0.724 MeV.}
\label{table:art2 30}
\end{table}

\renewcommand{\multirowsetup}{\centering}
\begin{table}[!h]
\centering
\begin{tabular}{cc|c|c|c|c|c|c|}
  \multicolumn{1}{c}{} & \multicolumn{1}{c}{} & 
  \multicolumn{6}{c}{$\sigma_k${\tiny $\left(MF+SV+PV\right)$}} \\
  \cline{3-8}
  \multicolumn{1}{c}{} & \multicolumn{1}{c|}{$\sigma_k${\tiny $\left(MF\right)$}} & 
    \multicolumn{1}{c}{$(a)$} & \multicolumn{1}{c|}{$(b)$} &
    \multicolumn{1}{c}{$(c)$} & \multicolumn{1}{c|}{$(d)$} &
    \multicolumn{1}{c}{$(e)$} & \multicolumn{1}{c|}{$(f)$} \\  \hline
  \multicolumn{1}{|c|}{\bf O} & 0.430 & 0.436	&	0.442	&	0.436	&	0.442	&	0.436	&	0.442	\\
  \multicolumn{1}{|c|}{\bf Si} & 0.603 & 1.113	&	1.141	&	0.939	&	0.973	&	0.939	&	0.973	\\
  \multicolumn{1}{|c|}{\bf S}  & 0.688 & 0.831	&	0.779	&	0.803	&	0.768	&	0.803	&	0.768	\\
  \multicolumn{1}{|c|}{\bf Ar} & 0.871 & 1.047	&	1.066	&	0.639	&	0.658	&	0.639	&	0.658	\\
  \multicolumn{1}{|c|}{\bf Ca} & 0.959 & 0.308	&	0.212	&	0.941	&	0.818	&	0.308	&	0.212	\\
  \multicolumn{1}{|c|}{\bf Ti} & 0.741 & 0.587	&	0.460	&	0.621	&	0.492	&	0.621	&	0.492	\\
  \multicolumn{1}{|c|}{\bf Fe} & 0.358 & 0.281	&	0.281	&	0.350	&	0.350	&	0.350	&	0.350	\\
  \multicolumn{1}{|c|}{\bf Se} & 0.136 & 0.703	&	0.789	&	0.250	&	0.161	&	0.250	&	0.161	\\
  \multicolumn{1}{|c|}{\bf Kr} & 0.296 & 0.649	&	0.776	&	0.298	&	0.319	&	0.298	&	0.319	\\
  \multicolumn{1}{|c|}{\bf Sr} & 0.548 & 0.572	&	0.718	&	0.519	&	0.550	&	0.519	&	0.550	\\
  \multicolumn{1}{|c|}{\bf Zr} & 0.535 & 0.673	&	0.814	&	0.461	&	0.502	&	0.461	&	0.502	\\
  \multicolumn{1}{|c|}{\bf Mo} & 0.352 & 0.332	&	0.522	&	0.208	&	0.245	&	0.208	&	0.245	\\
  \multicolumn{1}{|c|}{\bf Sn} & 0.498 & 0.621	&	0.525	&	0.473	&	0.436	&	0.473	&	0.436	\\
  \multicolumn{1}{|c|}{\bf Te} & 0.211 & 0.413	&	0.436	&	0.258	&	0.235	&	0.258	&	0.235	\\
  \multicolumn{1}{|c|}{\bf Xe} & 0.333 & 0.596	&	0.685	&	0.301	&	0.383	&	0.301	&	0.383	\\
  \multicolumn{1}{|c|}{\bf Ba} & 0.373 & 0.674	&	0.770	&	0.354	&	0.455	&	0.354	&	0.455	\\
  \multicolumn{1}{|c|}{\bf Ce} & 0.612 & 0.667	&	0.746	&	0.407	&	0.485	&	0.407	&	0.485	\\
  \multicolumn{1}{|c|}{\bf Sm} & 0.543 & 0.573	&	0.655	&	0.333	&	0.413	&	0.333	&	0.413	\\
  \multicolumn{1}{|c|}{\bf Hg} & 0.270 & 0.663	&	0.667	&	0.416	&	0.422	&	0.416	&	0.422	\\
  \multicolumn{1}{|c|}{\bf Pb} & 0.518 & 0.296	&	0.370	&	0.417	&	0.498	&	0.296	&	0.514	\\
  \multicolumn{1}{|c|}{\bf Po} & 0.359 & 0.359	&	0.295	&	0.416	&	0.344	&	0.416	&	0.385	\\ \hline
  \multicolumn{1}{|c|}{$\sigma_{total}$}  & \multicolumn{1}{|c}{\bf 0.532} 
        & \multicolumn{1}{|c}{\bf 0.623}	& \multicolumn{1}{c}{\bf 0.657}	& \multicolumn{1}{|c}{\bf 0.502}
  	& \multicolumn{1}{c}{\bf 0.497}	& \multicolumn{1}{|c}{\bf 0.464}	& \multicolumn{1}{c|}{\bf 	0.472}	\\ \hline
	\end{tabular}
\caption{Rms deviations $\sigma_k$ of calculated ground state energy plus correlation energies
from experimental values are shown.
These deviations are calculated with respect to the average deviation for each isotopic chain (see eqs. \ref{eq:art2 10} and
\ref{eq:art2 20}). The results in the different columns are obtained including different correlations, according
to the labels reported in Table III. The results reported in the fourth column (labeled (c)) 
have already been listed in the last column of Table I. All numbers are in MeV.}
\label{table:art2 50}
\end{table}

\section{Summary and conclusions}

We have performed a systematic analysis of the contribution of medium
polarization effects associated with collective surface and pairing 
vibrations on  the binding energies of even-even spherical nuclei.  
We have started from a HF+BCS mean field calculation based on a Skyrme-type
interaction, obtaining a rms deviation of 0.724 MeV with respect to experimental
masses. Then we have added the correlation energy associated with low-lying
surface and pairing vibrational modes (calculated in the QRPA and RPA approximation
respectively)
which typically increases the calculated binding energies by a few MeV.
Subsequently we performed a linear refit of the parameters of the 
interaction in order to obtain the best agreeement with the experiment.
Although this refit procedure had some difficulties in minimizing the rms deviation
of binding energies (some isotopic chains turned out to be
systematically overbound), we obtained in most isotopic chains 
a flattening of the deviations of the binding energies from the experiment.
Adding the fluctuations of the correlation energies, losing the information
related to their absolute value, the linear refit
was able to find a new set of Skyrme parameters which gives a rms
deviation of 0.691 MeV (that is, a reduction of about 4.5\% from the mean field treatment).

In order to improve the results achieved, we tried to better study 
some critical chains. For example the calcium isotopes, whose deviations
from experimental binding energies turned out to be insensitive to the addition
of the correlation energies considered, and which contribute in an 
important way to the total rmsd.
We have been able to improve the description of calcium and lead 
isotopes including the correlations associated with quadrupole pairing vibrations.
This reduced the rms deviation for all 121 isotopes to 0.667 MeV
(corresponding to  about 8\% reduction with respect to the MF rmsd).
The reduction obtained is not strong, but it is sufficient to justify 
the effort made in going beyond the mean field approach. The improvement gained by 
the specific study performed on Ca and Pb isotopes can suggest that
studying certain chains in more detail can allow us to learn something
that could be useful to build a general prescription.

Concerning surface vibrations correlations, 
we also compared our results with what was obtained by Bender et al. in ref. \cite{Ben.ea:2005}
with the projection on good angular momentum and obtained a very good agreement, 
pointing to the question of how much the details of the calculation count.

Our results show that, for the set of isotopes studied, the degrees of freedom we investigated can help us in
improving the mass formula built on a self-consistent mean field. 
However, our investigation was limited by the adopted refitting procedure, which
made a perturbative survey of the parameters space, tying the results obtained to the starting
Skyrme parametrization. Further investigations should adopt a more general refitting procedure,
exploring a larger region of the parameter space. Doing such a costly search for a better
minimum of the rms norm would also imply that at every step one should re-calculate the mean field binding energies
and also the correlation energies, because a larger change of the parameters implies that these correlation energies
can show a no more negligible variation. 

We conclude emphasizing that the aim of this work is not to achieve a mass formula with a very low rms deviation
from experimental findings, but to investigate the effects of some degrees of freedom of the nuclear system 
which have to be taken into account (namely, surface and pairing vibrations).

\section{Acknowledgements}
S.B. thanks G.F.Bertsch for the fruitful discussions at the Insitute of Nuclear Theory at Washington University, 
B.Sabbey for the helpful support about the refitting procedure and M.Pearson for his collaboration
about the mean field procedure.

\bibliographystyle{apsrev}

\begin{thebibliography}{99}
\bibitem{mat.ele} For pairing vibrations of multipolarity $\lambda^{\pi}$ for a nucleus with A nucleons 
we fitted the experimental values of the expression 
$2B_{A \pm 1}-B_{A \pm 2}-B_{A} + \Delta E^{exc} + \Delta E^{s.p.}$ for the pairing addition and removal mode, where
the $B$ are the binding energies,
$\Delta E^{exc}$ is the excitation energy of the lowest $\lambda^{\pi}$ state in the $A \pm 2$ nucleus 
and $\Delta E^{s.p.}$ is the energetic jump that the two valence nucleons have to make in order
to couple to angular momentum $\lambda$ ($\Delta E^{exc}$ and $\Delta E^{s.p.}$ are equal to zero for monopole 
($\lambda=0$) pairing vibration in even-even nuclei).
\bibitem{note:BE}One has to consider the electronic contribution to atomic binding energy.
\end{thebibliography}

\end{document}